\documentclass[usenatbib]{mn2e}
\usepackage{graphicx,times}
\usepackage{natbib}
\title[$f(R)$ theories constrained from lensing]{$f(R)$ gravity
theories in the Palatini Formalism constrained from strong lensing}
\author[Yang and Chen]{Xin-Juan Yang$^{1}$\thanks{E-mail:yxj@bao.ac.cn} and
Da-Ming
Chen$^{1}$\thanks{E-mail:cdm@bao.ac.cn}\\
$^{1}$National Astronomical Observatories,Chinese Academy of
Sciences,Beijing 100012,China}
\begin {document}
\maketitle
\begin{abstract}
$f(R)$ gravity, capable of driving the late-time acceleration of the
universe, is emerging as a promising alternative to dark energy.
Various $f(R)$ gravity models have been intensively tested against
probes of the expansion history, including type Ia supernovae
(SNIa), the cosmic microwave background (CMB) and baryon acoustic
oscillations (BAO). In this paper we propose to use the statistical
lens sample from Sloan Digital Sky Survey Quasar Lens Search Data
Release 3 (SQLS DR3) to constrain $f(R)$ gravity models. This sample
can probe the expansion history up to $z\sim2.2$, higher than what
probed by current SNIa and BAO data. We adopt a typical
parameterization of the form $f(R)=R-\alpha
H^2_0(-\frac{R}{H^2_0})^\beta$ with $\alpha$ and $\beta$ constants.
For $\beta=0$ ($\Lambda$CDM), we obtain the best-fit value of the
parameter $\alpha=-4.193$, for which the 95\% confidence interval
that is [-4.633, -3.754]. This best-fit value of $\alpha$
corresponds to the matter density parameter $\Omega_{m0}=0.301$,
consistent with constraints from other probes. Allowing $\beta$ to
be free, the best-fit parameters are $(\alpha, \beta)=(-3.777,
0.06195)$. Consequently, we give $\Omega_{m0}=0.285$ and the
deceleration parameter $q_0=-0.544$. At the 95\% confidence level,
$\alpha$ and $\beta$ are constrained to [-4.67, -2.89] and [-0.078,
0.202] respectively. Clearly, given the currently limited sample
size, we can only constrain $\beta$ within the accuracy of
$\Delta\beta\sim 0.1$ and thus can not distinguish between
$\Lambda$CDM and $f(R)$ gravity with high significance, and
actually, the former lies in the 68\% confidence contour. We expect
that the extension of the SQLS DR3 lens sample to the SDSS DR5 and
SDSS-II will make constraints on the model more stringent.

\end {abstract}
\section{introduction}
In modern cosmology, one of the most striking discoveries is that
our expanding universe is undergoing a phase of acceleration. The
key observational results that support this discovery are: the
luminosity-redshift relationship from SNIa surveys
\citep{Astier2006, Perlmutter1999, Riess1998, Riess2004, Riess2007},
the CMB anisotropy spectrum \citep{de Bernardis2000, Spergel2003,
Spergel2007, Hinshaw2007}, the large scale structure from galaxy
redshift surveys \citep{Cole2005, Seljak2005, Tegmark2004a} and BAO
\citep{Eisenstein2005} . With the combination of all these
observational results, one gets the standard $\Lambda$CDM cosmology
as: $\Omega_K=0,\Omega_M=0.27,\Omega_\Lambda=0.73$.  Amongst the
matter content baryonic matter amounts to only $4\%$, other $23\%$
is the so-called dark matter (DM), and the rest component of $73\%$
dominate the universe that is often referred to as dark energy (DE), which is a
negative-pressure ideal fluid smoothly permeated the universe. It is
worth noting that, notwithstanding the standard cosmology is very
well in agreement with the astrophysical data, the nature of the
dark matter and dark energy remains mystery in the modern Cosmology
and Physics, so that there are so many theoretical proposals to
account for them on the ground. Possible models to explain the
acceleration include a classical cosmology constant
\citep{Weinberg1989, Carroll1992}, Chaplygin gas
\citep{kamenshchik2001,zhu2004,wuyu2007a,wuyu2007b}, a wide variety
of scalar-field models such as Quintessence
\cite[e.g.,][]{Caldwell1998,wu2008}, K-essence
\citep{Armendariz-Picon2000, Armendariz-Picon2001}, Phantom field
\citep{Caldwell2002, Dabrowski2003} and so on.

Except above models, there are several popular alternative ideas for
the accelerating universe by modifying general relativity rather
than resorting to some kinds of exotic fluid. It is a natural
consequence to modify the theory of gravity, after the current
observations mightily suggesting the failure of General Relativity (GR) as a large
cosmological scale gravity theory. Actually, on the right side of
Einstein's field equation, the energy momentum tensor is related to
the matter content, and on the left side,the Einstein tensor
consists of pure geometrical terms, which can be modified to account
for DM and DE phenomena \citep{Copeland2006}. Accordingly, both of
accelerating behaviour (DE) and dynamical phenomena (DM) can be
interpreted as curvature effects. Recently, a lot of people attempt
to modify the General Relativity in different ways and propose
various modified gravity theories to address the DM or DE phenomenon
or both. Amongst them, $f(R)$ gravity theories, where $R$ is
curvature scalar, have attracted much attention, which are put
forward to account for DE by adopting an arbitrary function of R in
Einstein-Hilbert Lagrangian, the so-called $f(R)$ term, instead of
$R$ in traditional General Relativity. Obviously, an important
reason for this interest is that $f(R)$ term in types of negative as
well as positive powers of R can produce the inflation at early
times and now observed acceleration phase at late times, following
the well-known sequence of universal evolution
\citep{Nojiri2003, Nojiri2007, Sotiriou2006-2, Capozziello2006-1, Vollick2003, Chiba2003,
Erickcek2006, Navarro2007, Olmo2005, Olmo2007, Amendola2007, Li
B.2007-1, Li B.2006}. Usually, two possibilities referring to as the
metric approach and the Palatini approach can derive the Einstein
field equation from the Hilbert action. By comparison, the metric
approach assumes that the equation has only one variable with
respect to metric where the affine connection is the function of
metric, while in the Palatini approach the metric and the connection
are treated to be independent of each other as two variables. When
$f(R)$ is in form of linear function, both approaches lead to the
same results.  Once adding any non-linear term to the Hilbert
action, however, the two methods would produce enormous differences.
The reasons for the Palatini approach seeming appealing compared
with another one are that, on one hand, the metric approach leads to
the fourth order field equations, which is difficult to solve
analytically, whereas the Palatini approach results in the two order
field equations. On the other hand, although  $f(R)$ theories via
metric approach can give some interesting and successful results,
some of them suffer from certain fatal defects. For example, they
cannot pass the solar system test, have instabilities, have
incorrect Newtonian limit and cannot totally describe all epoches of
the universe.

Here we adopt the $f(R)$ gravity theory within the Palatini approach
which can avoid the above mentioned problems. In this framework, the
form $f(R)=R+\alpha(-R)^\beta$ is chosen so that it can pass the
solar system test and has the correct Newtonian limit
\citep{Sotiriou2006-3}, and can explain the late accelerating phase
in the universe \citep{Fay2007, Capozziello2005, Sotiriou2006-1}.
Furthermore, it is important to go beyond the quantitative studying
and test these theories using the observations. Recently constraint
from data on above type of $f(R)$ theory is intensively discussed by
many authors. Among them, in these papers \citep{Santos2008, Fay2007,
Borowiec2006, Amarzguioui2006, Sotiriou2006-1}, the CMB shift
parameter, supernovae Ia surveys data and baryon acoutic
oscillations were combined to constrain the parameters and, in
particular, they give $\beta\sim 10^{-1}$. In a previous work,
\citet{Koivisto2006} used the matter power spectrum from the SDSS to
get further restriction $\beta\sim 10^{-5}$, which reduced allowed
parameter space to a tiny around the $\Lambda$CDM cosmology.
Recently, \citet{Li B.2007-2} jointed the WMAP, supernovae Legacy
Survey (SNLS) data and Sloan Digital Sky Survey (SDSS) data to
tighten the parameter up to $\beta\sim 10^{-6}$, which made this
model hard to distinguish from the standard one, where the parameter
$\beta$ is zero. So far, there has been no attempt to use the strong
lensing observation to test the $f(R)$ gravity theories in the
Palatini formalism.

Nowadays, with more and more lens surveys available
to enlarge the statistical lensing samples,
gravitational lensing has developed into a powerful tool to study a
host of important subjects on different scales in astrophysics,
from stars to galaxies and clusters, further, to the large
structure of the universe. Since the gravitational lensing phenomena
involve the source information, two dimensional mass distribution of
the lens and the geometry of the universe, it is useful to not only
infer the distant source properties far below the resolution limit
or sensitivity limit of current observations and offer an ideal way
to probe the mass distribution of the universe, but also constrain
the parameters of cosmological models \citep{Copeland2006, Wu1996}.
So far the large lens surveys have built both radio (the
Cosmic-Lens All Sky Survey, CLASS) and optical (SQLS DR3) lens samples. The
CLASS forms a well-defined lens
sample containing 13 lenses from 8958 radio sources with
image separations of $0.3^{''}<\theta<3^{''}$ and the i-band
flux ratio $q\leq10$ (bright to faint) \citep{Browne2002}. Meanwhile,
the SQLS constructs a statistical sample of 11 lensed
quasars from 22683 optical quasars, of which the range of redshift is
$0.6<z<2.2$ and the apparent magnitude is brighter than $i=19.1$
\citep{Inada2008}. As compared with CLASS, the latter sample has
larger image separations ($1^{''}<\theta<20^{''}$) and smaller flux
ratio limit ($q_c=10^{-0.5}$, faint to bright). Many authors use the
strong gravitational lensing statistics to study DE,
including the equation of state of DE \citep{Chae2002,
Chen2004, Oguri2008} and the test for the modified gravity theories as
alternatives to DE
\citep{Zhu2008}.

The main goal of this paper is to investigate the constraints of
strong lensing observation on the parameters of $f(R)$ theory in the
Palatini approach using the SQLS statistical sample. But there has a
big question that how the non-linear terms in $f(R)$ gravity
theories contribute to the light-bending and correct the Einstein
deflection angle. If this influence is large enough, it is
inevitable that all the models built on GR to compute the lensing
statistical probability must be changed. But this question is
starting to be solved, here, we still use the same way as that based
on GR to calculate the lensing probability because of the following
considerations. Firstly, the $f(R)$ gravity theories are applied to
explain the nature of DE, but as far as we know it, DE smoothly
fills the space to power the acceleration expansion as the same way
in everywhere of the universe and make the universe to be flat.
\citet{Dave2002} showed that DE does not cluster on scales less than
100Mpc. Naturally, as an alternative to DE, any modified gravity
theory, like $f(R)$ non-linear term, should not affect the
gravitational potential distribution of dark halos as the virialized
systems such as galaxies and clusters, and in particular, should
have not detectable contribution to the light-bending if they are some successful models.
Moreover, it is worth stressing that how the modifying the
lagrangian of the gravitational field affects the standard theory of
lensing (built on GR) is not well investigated. Recently, theories
with $f(R)\propto R^n$ were been investigated within the metric
approach for a point-like lens and results were given that the $R^n$
modified gravity signatures possibly to be detected through a
careful examination of galactic microlensing
\citep{Capozziello2006-2}. And \citet{Zhang2007} derived the
complete set of the linearized field equations of the two Newtonian
potentials $\phi$ and $\psi$ in the metric formalism, and predicted
that, for some $f(R)$ gravity models, the gravitational lensing was
virtually identical to that based on GR, under the environments of
galaxies and clusters. While, under the Palatini approach, the paper
\citep{Ruggiero2008} is a first step to evaluate some effects of the
non linearity of the gravity Lagrangian on lensing phenomenology. As
expected, for a spherically symmetrical point-like model, the
estimated results suggest that the effects of $f(R)$ are confined
around a cosmological scale, and hence, they have no effects on
smaller scales such as our Galaxy. For galaxies, the
weak-field approximation is valid. The standard lensing theory
(based on GR) is on the basis of this assumption. In GR, the
deflection angle for a point mass can be easily obtained. For the
weak-field assumption, the field equations of GR can be linearized.
Therefore, we can first divide the whole galaxy into small elements,
and each mass element can be treated as point mass, then the
deflection angle of the general galactic mass distribution models
can be obtained by the sum (or integration) of the deflections due
to the individual mass components \citep{Kochanek2004}. In $f(R)$
gravity, if the gravitational field is weak, we are able to perturb
the General Einstein field equations (based on $f(R)$ gravity
theories) and simplify them to linear equations. In this case, the
superimposition principle works, so we can get that the field
equations which describe the point-like objects can also be used to
describe the ensemble of mass points. That is to say,  both of the
mass point and the ensemble of mass points have field equations in
the same formalism. So we can safely extrapolate the conclusion for
point-like lenses, that standard lensing theory is a good
approximation in the case of $f(R)$ gravity theories, to
galaxy-scale lenses. However, there are no works to prove it through
mathematical process, and further theoretical studies are needed to
test the $f(R)$ gravity theories in the Palatini formalism. Under these situations, we consider that the standard theory of lensing is still correct in the caseof $f(R)$ theory we used.

According to astronomical observations, dynamical analysis
and numerical simulations, there are three kinds of popular mass
density profile of dark halos as a lens model in the standard
lensing theory, namely, the singular isothermal sphere (SIS), the
Navarro-Frenk-White (NFW) and the generalized NFW (GNFW) profile.
They can reasonably reproduce the results of strong lensing survey
\citep{Li2002, Li2003, Sarbu2001}. As been well known, the lensing
probability is very sensitive to the density profile of lenses.
Moreover, the lensing is determined almost entirely by the fraction
of the halo mass that is contained within a fiducial radius that is
$\sim\%4$ of the virial radius. Since in the inner regions the slope
of SIS density is steeper than that of NFW density, for the multiple
image separations of $\Delta\theta<5{''}$, the lensing probability for
NFW halos is lower than the corresponding probability for SIS halos
by about 3 orders of magnitude \citep{Li2002, Li2003, Keeton2001-2,
Wyithe2001}. The cooling mass dividing galaxies and clusters for
transition from SIS to NFW is $M_c\sim10^{13}M_\odot$. The SQLS DR3
has 11 multi-imaged lens systems, including ten galaxy-scale lenses
and one cluster-scale lens with large image separation ($14.62{''}$)
\citep{Inada2008}. It is the first lens sample that contains both
galaxy-scale lenses and cluster-scale lens. In lensing statistics,
if we consider lensing halos with all mass scales, from galaxies to
clusters,  then the effects of substructures \citep{Oguri2006} and
baryon infall \citep{kochanek2001,keeton2001-1} cannot be neglected,
in particular to galaxy groups, for which neither SIS nor NFW can be
applied. For early-type galaxies, which dominate the galactic strong
lensing, SIS is a good model of lenses, in particular, it results
from the original NFW-like halos through the baryon infall effects
\citep{rusin2005,koopmans2006}. Moreover, the non-spherical lens
profiles do not significantly affect the lensing cross section
\citep{Oguri2005, Huterer2005, Kochanek1996}. Therefore, in this
paper, we use the SIS model to compute the strong lensing
probability for galaxies. To avoid the complexity from large mass
scales mentioned above, we consider only galaxy-scale lenses in the
sample, and thus compute the differential lensing probability rather
than the usual integrated lensing probability to match the SQLS DR3
sample (containing 10 galaxy-scale lenses).  A more realistic
lensing model including SIS and NFW to fit the complete SQLS sample
and comparing to the results of this paper is our future work. As
comparison, \citet{Oguri2008} adds two additional cuts to select a 7
lensed quasar sub-sample from the DR3 statistical lens sample for
constraining the parameter $\omega$ of state of dark energy.

The rest of the paper is organized as follows: In section 2 we
outline the $f(R)$ theories in the Palatini approach, including the
cosmological dynamical equation and parameters with the FRW universe
setting.  In section 3 we present the differential lensing
probability based on the $f(R)$ gravity theories with SIS model. In
section 4, we discuss the specific analytic function $f(R)=R-\alpha
H^2_0(-\frac{R}{H^2_0})^\beta$ used in this paper, give the
corresponding cosmological evolution behaviour and investigate the
observational constraint on our $f(R)$ theory arising from SQLS DR3
statistical lens sample. Finally, we give discussion and conclusions
in section 5.

\section[]{the $f(R)$ gravity theories and corresponding cosmology parameters}

\subsection{The generalized Einstein Equations}

The starting point of the $f(R)$ theories in the Palatini approach
is Einstein-Hilbert action, whose formalism is:
\begin{equation}
S=-\frac{1}{2\kappa}\int{d^4x\sqrt{-g}f(R)}+S_M,
\label{action}
\end{equation}
where $\kappa=8\pi G$, light velocity $c=1$ and $S_M$ is the matter
action which is a functional of metric $g_{\mu\nu}$ and the matter
fields $\psi_M$. As mentioned previously, the metric and
the affine connection are two independent quantities in the Palatini
approach, so the Ricci scalar is defined by two variables, the
metric and the connection \cite[see][for details]{Vollick2003}:
\begin{equation}
R=g^{\mu\nu}\hat{R}_{\mu\nu},
\end{equation}
in which
\begin{equation}
\hat{R}_{\mu\nu}={\hat{\Gamma}^\alpha}_{\mu\nu,\alpha}-{\hat{\Gamma}^\alpha}_{
\mu\alpha,\nu}+{\hat{\Gamma}^\alpha}_{\alpha\lambda}{\hat{\Gamma}^\lambda}_{
\mu\nu}-{\hat{\Gamma}^\alpha}_{\mu\lambda}{\hat{\Gamma}^\lambda}_{\alpha\nu},
\end{equation}
where R is negative. From Eq.(\ref{action}), we derive the generalized Einstein
equations.
Varying with respect to metric $g_\mu\nu$, we get one equation:
\begin{equation}
f'(R)\hat{R}_{\mu\nu}-\frac{1}{2}g_{\mu\nu}f(R)=-\kappa
T_{\mu\nu},
\label{equation}
\end{equation}
where $f'(R)=df/dR$ and the energy momentum tensor $T_{\mu\nu}$ is
given as
\begin{equation}
T_{\mu\nu}=-\frac{2}{\sqrt{-g}}\frac{\delta S_M}{\delta
g^{\mu\nu}}.
\end{equation}
Taking the trace of Eq.(\ref{equation}) gives the so-called structural equation
and it controls the solutions of Eq.(\ref{equation}).
\begin{equation}
Rf'(R)-2f(R)=-\kappa T.
\label{structure}
\end{equation}
 Varying with respect to the connection
${\hat{\Gamma^\lambda}_{\mu\nu}}$ gives another equation
\begin{equation}
\hat{\nabla}_\alpha[f'(R)\sqrt{-g}g^{\mu\nu}]=0,
\end{equation}
where $\nabla$ denotes the covariant derivative with respect to the
affine connection. From this equation it is found that the new
metric $h_{\mu\nu}$, which can describe the affine connection as the
Levi-Civita connections, is conformal to $g_{\mu\nu}$,
\begin{equation}
h_{\mu\nu}=f'(R)g_{\mu\nu}.
\end{equation}
Finally combined above equations, we could easily get the relation
between $\hat{R}_{\mu\nu}$ and ${R}_{\mu\nu}$ which is the Ricci
tensor associated with the metric $g_{\mu\nu}$.
\begin{equation}
\hat{R}_{\mu\nu}=R_{\mu\nu}-\frac{3}{2}\frac{\nabla_\mu f'\nabla_\nu
f'}{f'^2}+\frac{\nabla_\mu\nabla_\nu
f'}{f'}+\frac{1}{2}g_{\mu\nu}\frac{\nabla^\mu\nabla_\mu f'}{f'}.
\label{ricci}
\end{equation}
Note that the covariant derivative $\nabla$ we refer to is
associated with the Levi-Civita connection of metric $g_{\mu\nu}$.

\subsection{The Background Cosmology}
In this subsection, we shall make a detailed study of the
cosmological viability of the model based on $f(R)$ theories in Palatini
formalism. It is well-known that WMAP's data is a strong evidence to support
a flat universe, so we choose a spatially flat FRW
(Friedmann-Robertson-Walker) metric to describe our background
universe. The metric takes the standard form:
\begin{equation}
ds^2=-dt^2+a(t)^2\delta_{ij}dx^idx^j             (i,j=1,2,3),
\end{equation}
where $a(t)$ is the scale factor. As usual, we assume a perfect fluid
energy-momentum tensor
${T_\mu}^\nu=diag(\rho,p,p,p)$. By contracting Eq.(\ref{ricci}) we get the
generalized Friedmann equation:
\begin{equation}
(H+\frac{1}{2}\frac{\dot{f'}}{f'})^2=\frac{1}{6}\frac{\kappa(\rho+3p)}{f'}-\frac
{1}{6}\frac{f}{f'}.
\label{friedmann}
\end{equation}
In this paper we consider a matter dominated universe, so the
constant equation of state is $p=\omega\rho(\omega=0)$ and the
relation between the matter density and scalar factor is
$\rho=\rho_0R^{-3(\omega+1)}(\omega=0)$. Using
Eq.(\ref{structure}),  we can obtain:
\begin{equation}
a(R)=(\kappa\rho_{m0})^\frac{1}{3}(Rf'-2f)^{-\frac{1}{3}},
\label{ar}
\end{equation}
where we have chosen $a_0=1$. On the other hand, from Eq.(\ref{structure}) and
the
conservation of energy $T_{\mu\nu;\lambda}=0$, we have
as:
\begin{equation}
\dot{R}=-\frac{3H\rho_M}{Rf''(R)-f'(R)}.
\label{dotr}
\end{equation}
Using Eqs.(\ref{structure}), (\ref{friedmann}) and (\ref{dotr}) we get the
Friedmann equation $H(R)$ in
the form of R,
\begin{equation}
H^2(R)=\frac{1}{6f'}\frac{Rf'-3f}{(1-\frac{3}{2}\frac{f''(Rf'-2f)}{f'(Rf''-f')}
)^2}.
\label{h2r}
\end{equation}
Combing Eqs.(\ref{ar}) and (\ref{h2r}) we can know the whole expansion history
that is determined by $H(a)$, for any specific expression of $f(R)$.

Let us now consider the cosmological distance based on the
given model. From the explicit expression for the Hubble parameter
$H(R)$ and the relation
between z and $a(R)$: $1+z=a^{-1}(R)$, it is convenient to rewrite
the proper distance, luminosity distance, angular diameter distance
and the deceleration parameter in terms of R. The proper distance and luminosity
distance to the object at redshift $z$ are
\begin{eqnarray}
D^P(z) &=&\int^z_0{\frac{dz}{(1+z)H(z)}} \\ \nonumber
&=&\frac{1}{3}\int^{R_z}_{R_0}{\frac{Rf''-f'}{Rf'-2f}\frac{dR}{H(R)}}=D^P(R),
\end{eqnarray}
\begin{eqnarray}
D^L(z)&=&(1+z)\int^z_0{\frac{dz}{H(z)}} \\ \nonumber
&=&\frac{1}{3}(k\rho_0)^{-\frac{2}{3}}(Rf'-2f)^{\frac{1}{3}}\int^{R_z}_{R_0}{
\frac{Rf''-f'}{(Rf'-2f)^\frac{2}{3}}\frac{dR}{H(R)}}
\\ \nonumber
&=&D^L(R).
\end{eqnarray}
The angular diameter distance from an object at red-shift $z_1$ to
an object at red-shift $z_2$ is
\begin{eqnarray}
D^A(z_1,z_2)&=&\frac{1}{1+z_2}\int^{z_2}_{z_1}{\frac{dz}{H(z)}}
\\ \nonumber
&=&\frac{1}{3}(Rf'-2f)^{-\frac{1}{3}}\int^{R_{z_2}}_{R_{z_1}}{\frac{Rf''-f'}{
(Rf'-2f)^\frac{2}{3}}\frac{dR}{H(R)}}
\\ \nonumber
&=&D^A(R_1,R_2).
\end{eqnarray}
The deceleration parameter is
\begin{equation}
q(t)=-\frac{a(t)\ddot{a}(t)}{\dot{a}^2(t)}=-\left(1+\frac{H'(R)a(R)}{H(R)a'(R)}
\right)=q(R),
\end{equation}
when $q>0$ the universe is in early-time
matter-dominated phase and  $q<0$ the universe is in dark energy dominated
late-time acceleration phase.

\section{The lensing probability}

\subsection{the SIS lens}
As mentioned, in this paper, we assume that the
$f(R)$ theories have no effect on the gravitational lensing and
still model the early-type galactic lens as SIS as in GR, but these formulae are
used in terms of R. The
density profile is
\begin{equation}
\rho(r)=\frac{\sigma^2}{2\pi G}\frac{1}{r^2},
\end{equation}
where $\sigma$ is the velocity dispersion and $r$ is the distance
from the galaxy center.  From the geometrical relations between the
image position $\theta$, source position $\beta$ and lens, one gets the
lensing equation:
\begin{equation}
\beta=\theta-\frac{D^A_{LS}}{D^A_S}\alpha.
\end{equation}
For a SIS lens, the deflection angle is $\alpha=4\pi(\sigma/c)^2$,
which is independent of the impact parameter. The
angular Einstein radius is defined as
\begin{equation}
\theta_E=4\pi(\frac{\sigma}{c})^2\frac{D^A_{LS}}{D^A_S}.
\label{radius}
\end{equation}
The sources are multiply imaged when
$\beta<\theta_E$ and a ring-like image occurs when
$\beta=0$. When the multiple images occur, from the lensing
equation, there are two images on opposite sides of the lens at
angular positions
\begin{equation}
\theta_\pm=\theta_E\pm\beta.
\end{equation}
The corresponding magnifications are
\begin{equation}
\mu_\pm=\frac{\theta_\pm}{\beta_\pm}\frac{d\theta_\pm}{d\beta_\pm}=\frac{
\theta_E\pm\beta_\pm}{\beta_\pm},
\end{equation}
the total magnification of the two images is
\begin{equation}
\mu_{tot}=\mu_++\mu_-=\frac{2\theta_E}{\beta}
\end{equation}
The magnification of an image is defined by the ratio between the
solid angles of the image and the source, namely, the flux density
ratio between the image and the source. Then the flux density ratio
q for bright-to-faint is the ratio of the corresponding absolute
values of the magnifications
\begin{equation}
q=|\frac{\mu_+}{\mu_-}|=\frac{\theta_E+\beta}{\theta_E-\beta}.
\label{q}
\end{equation}
Every lens survey has its own spectrum range and selection
functions, so there exists an explicit cut flux density ratio, less
than which, the survey couldnot identify the lenses. For example,
for SQLS, the ratio (faint to bright) is larger than
$q_c=10^{-0.5}$, which means that only sources with
$\beta<\beta_{max}<\theta_E$ can be detected, where
$\beta_{max}$ corresponds to $q_c$.
 The lensing cross section is
\begin{equation}
\sigma_{SIS}(\sigma)=16\pi^3(\frac{\sigma}{c})^4(\frac{D^A_{LS}D^A_L}{D^A_S})^2,
\label{sigma-sis}
\end{equation}
and the velocity to produce image separation $\Delta\theta$ is
\begin{equation}
\sigma_{\Delta\theta}=4.392\times10^{-4}(\frac{c}{\sigma_\star})(\frac{D^A_S}{
D^A_{LS}})^{\frac{1}{2}}{(\Delta\theta^{''})}^{\frac{1}{2}}.
\end{equation}

\subsection{the differential probability}
The differential probability for the source quasar at the redshift
$z_s$ (corresponding to $R_s$) lensed by foreground dark halos with
multiple image separations $\Delta\theta$ and flux density ratio
larger than $q_c$ is given by
\begin{eqnarray}
P(\Delta\theta,R_s,>q_c)=\frac{dP(>\Delta\theta,R_s,>q_c)}{d\Delta\theta}
\\ \nonumber
=\int^{R_s}_0dR\frac{dD^P(R)}{dR}[n(\sigma,R_z)a^{-3}(R)]\sigma_{real}(\sigma,
R_z,>q_c)\frac{d\sigma}{d{\Delta\theta}}
\end{eqnarray}
where $D^p$ is the proper distance from the observer to the lens.
$n(\sigma,R_z)$ and $n(\sigma,R_z)a^{-3}(R)$ are the comoving number
density and the physical number density of galaxies at $R_z$
(corresponding to redshift z) with dispersion between $\sigma$ and
$\sigma+d\sigma$, respectively. the cross section
$\sigma_{real}(\sigma,R_z,>q_c)$ is dependent on the flux density
ratio of multiple images ($q_c$) and take into account the
magnification bias (see below for details).

In the statistics for gravitational lensing, there are two
independent ways to get the mass function of virialized halos.
One is the generalized
Press-Schechter (PS) theory, and the other is Schechter luminosity
function. Since the PS
theory in the framework of $f(R)$ theories does not exist, we adopt
the latter. It is established that \citep{Maoz1993,Moller2006,Chae2007}, the
early-type galaxies
dominant strong lensing, the contribution of the late-type galaxies
is well neglected in particular when the image separations are
larger than $1''$, the reason is that the late-type galaxies have
larger rotation velocity while have smaller dispersion velocity, which in
turn, lead to smaller mean image separations. Moreover, in our used
SQLS lens sample all lens galaxies are early-type. Many previous studies
for lensing statistics applying the no-evolution of the velocity
function get appealing results to agree with the galaxy number
counts \citep{Im2002} and the redshift distribution of lens galaxies
\citep{Chae2003, Ofek2003}. Meanwhile, other studies \citep{Mao1994,
Mitchell2005, Rix1994} have investigated the effects of evolution of the
velocity
function on the lensing statistics, and they concluded that the
simple evolution does not significantly affect lensing statistics if all
galaxies are early-type. In our paper, we do not take
into account the evolution of the velocity function, instead, we use
the modified Schechter function \citep{Chae2007,
Chen2006, Oguri2008, Mitchell2005, Zhu2008}
\begin{equation}
\phi(\sigma)d\sigma=\phi_\star(\frac{\sigma}{\sigma_\star})^\alpha
\exp[-(\frac{\sigma}{\sigma_\star})^\beta]\frac{\beta}{\Gamma(\alpha/\beta)}\frac
{d\sigma}{\sigma},
\end{equation}
with $(\phi_{\star},\sigma_{\star},\alpha,\beta)=(0.008h^3\mbox{Mpc}^{-3},
161\mbox{kms}^{-1}, 2.32, 2.67)$,
which is derived by \citet{Choi2007} upon the latest much larger
SDSS DR3 data. Then the comoving number density is
$n(\sigma,R_z)=\phi(\sigma)$, and  $\sigma$ is in terms of $R_z$.

\begin{figure}
\includegraphics[width=84mm]{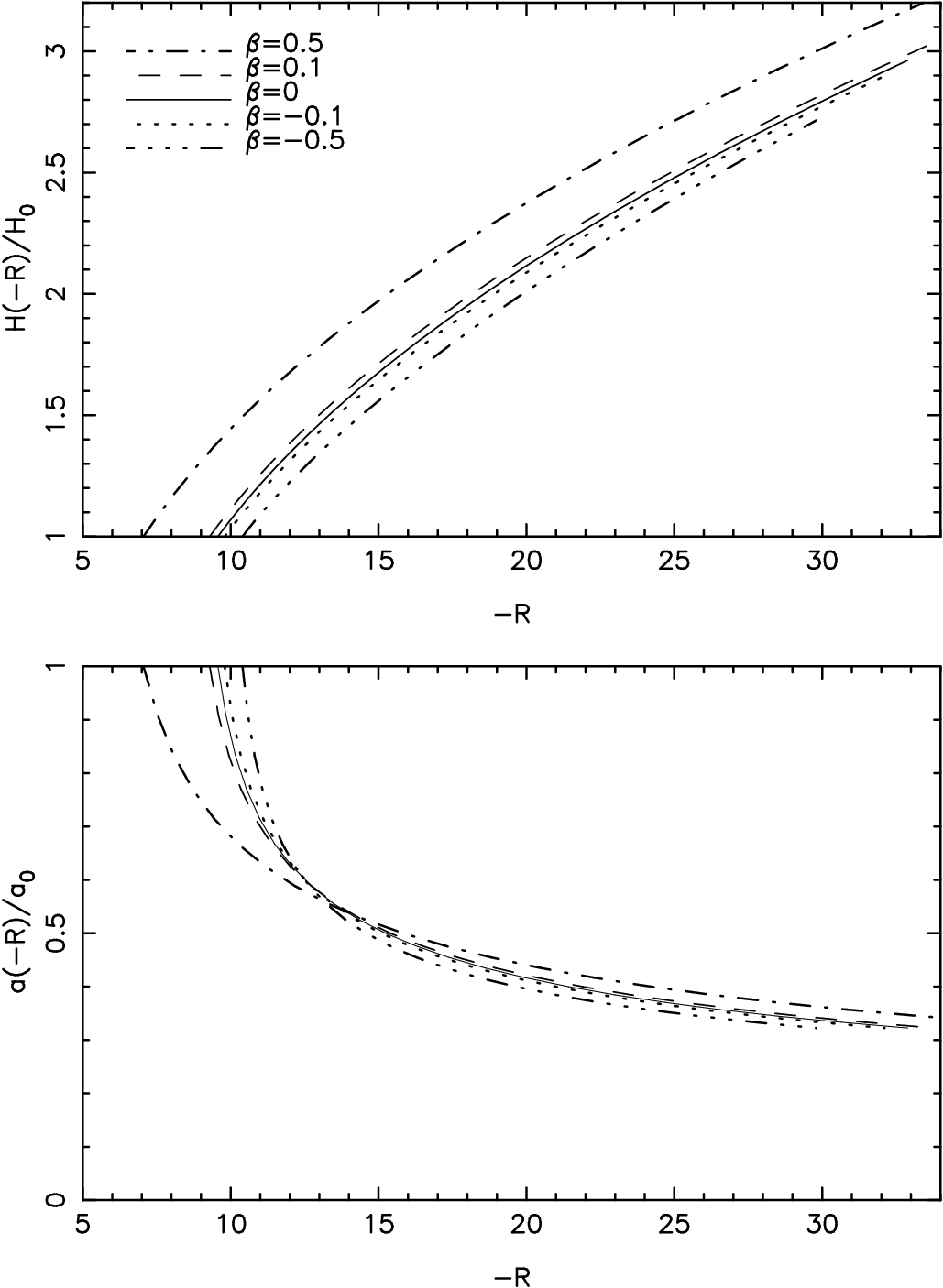}
\caption{The cosmology parameter as a function of R. The upper one
is the hubble parameter $H(-R)$. The other one is the scale factor
$a(-R)$.} \label{ahr}
\end{figure}

Lensing probabilities must be corrected with the magnification bias,
which explains the fact that intrinsically quasar sources
with flux below the flux-limited survey can appear by virtue
of lensing magnification. The cross section $\sigma_{real}(\sigma,R_z,>q_c)$
including the
 magnification bias parameter B is
defined as \citep{Oguri2008}
\begin{equation}
\sigma_{real}=\int{d\textbf{\textit{u}}\frac{\phi(L_{min}/\mu)}{\mu\phi(L_{min})}},
\label{sigma-real}
\end{equation}
here $d\textbf{\textit{u}}=D^2_s\beta d\beta d\omega (\omega\in(0,2\pi))$ is the
area element of the lens plane. The magnification ratio $\mu$ in the
best-fitted form is \citep{Oguri.etal.2006}
\begin{equation}
\mu=\bar{\mu}\mu_{tot}+(1-\bar{\mu})\mu_+,
\end{equation}
here
\begin{equation}
\bar{\mu}=\frac{1}{2}[1+\tanh(1.76-1.78\theta'')].
\end{equation}
By defining $t=\beta/\theta_E$ and using Eqs.(\ref{radius}), (\ref{q}),
(\ref{sigma-sis}) and (\ref{sigma-real}), we can rewrite the cross section as,
\begin{equation}
\sigma_{real}(\sigma,R_z,>q_c)=\sigma_{SIS}(R_s,\Delta\theta)B(R_s,\Delta\theta,
q_c),
\end{equation}
here the bias $B$ for the multiply imaged source at scalar $R_s$
is
\begin{equation}
B(R_s,\Delta\theta,q_c)=2\int^{t_c}_0{\frac{tdt}{\bar{\mu}\frac{2}{t}
+(1-\bar{\mu})(\frac{1}{t}-1)}\frac{\phi(L_{min}/\mu)}{\phi(L_{min})}},
\end{equation}
where $tc=(1-q_c)/(1+q_c)$ and $\phi(L_{min})$ is the quasar
differential luminosity functions at the luminosity limit $L_{min}$
of the survey, which is described by a typically double power-law
in terms of absolute magnitude in g-band at
$R_z$ by \citep{Hopkins2007, Li G.L.2007}
\begin{equation}
\phi(M_g)=\frac{\phi_\ast}{10^{0.4(1+\beta_h)(M_g-M^\ast_g)}
+10^{0.4(1+\beta_h)(M_g-M^\ast_g)}},
\end{equation}
and
\begin{eqnarray}
M^{\ast}_g(R_s)&=&M^{\ast}_g(R_0)-2.5[k_1(a^{-1}(R_s)-1) \nonumber \\
 & &+k_2(a^{-1}(R_s)-1)^2].
\end{eqnarray}

For the low redshift $(z<2.1)$ quasars, the parameters from the
2dF-SDSS Luminous Red Galaxy (LRG) and Quasi-stellar Object (QSO) Survey fitted in the SDSS g-band are
$(\phi_\ast, \beta_h, \beta_l, M^\ast_g(R_0), k_1,
k_2)=(1.83\times10^{-6}(h/0.7)^3Mpc^{-3}mag^{-1}, -3.31, -1.45,
-21.61+5log(h/0.7), 1.39, -0.29)$. By virtual of the relation between
luminosity and absolute magnitude $L\propto10^{-0.4(M-M^\ast)}$, we
have  $\phi(L/\mu)$ in terms of absolute magnitude. The SDSS
quasar survey have flux limit in i-band and the apparent
magnitude limit is $i_{max}=19.1$, but the parameters in  above
luminosity Equations are given in g-band, so we first convert the
apparent magnitude to the absolute magnitude, then compute the
corresponding absolute magnitude in g-band using the K-corrections
which is given by \citep{Richards2006}

\begin{equation}
M_{g}(R_0)=M_i(R_0)+2.5\alpha_{\nu}
\log\left(\frac{4670{\AA}}{7471{\AA}}\right)-0.187,
\end{equation}
where $\alpha_{\nu}=-0.5$.

From the differential lensing probability, we calculate the expected
number of lensed quasars from the SQLS sample, and compare to the
observational results in order to constrain the parameters in $f(R)$ models. The
results will
be given in the next section.

\section{numerical results and observational constraints}

Here, what we would like to investigate is what extent observations
of strong lensing allow deviations from the GR which is
corresponding to $f(R)=R$. For our purposes, we adopt the following
representation for $f(R)$ \citep{Amarzguioui2006}.
\begin{equation}
f(R)=R-\alpha H^2_0(-\frac{R}{H^2_0})^\beta,
\end{equation}
which is a possible candidate for the late-time cosmic accelerating
expansion found recently. Here, $H_0$ is a constant with dimension
Mpc$^{-1}$, which are introduced to make the $\alpha$ and $\beta$
dimensionless. Not all combinations of $\alpha$ and $\beta$ are
agreement with a flat universe with matter dominated era flowed by
an accelerated expansion today. We now consider the constraint on
these parameters. Firstly, at the early time of matter dominated
universe, the dark energy is ineffective and the universe is better
described by GR. Therefore, with $-R$ ($R$ is negative ) being more
and more larger, the modified Lagrangian should approach its GR
limit, in order to avoid confliction with early-time physics such as
Big Bang Nucleosynthesis (BBN) and CMB, and hence it is
straightforward to demand that $\beta<1$. Moreover, in the case of
$T=-\rho$, we must demand that the left side of Eq.(\ref{structure})
and the right of Eq.(\ref{h2r}) be always positive. In the special
case of $(\alpha,\beta)=(-4.38,0)$, the $\Lambda$CDM cosmology model
is recovered.

\begin{figure}
\centering
\includegraphics[angle=270,width=84mm]{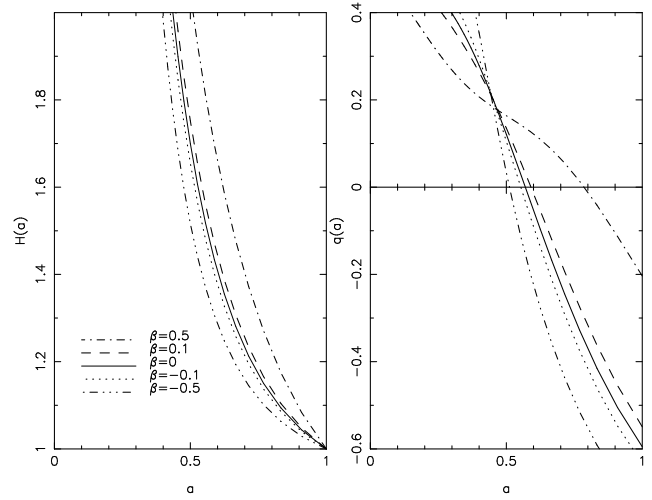}
\caption{The background dynamical behaviour based on the
$f(R)=R-\alpha(-R)^\beta$ model. The left panel is the hubble
parameter $H$ as a function of scale factor $a$. The right panel is
the deceleration parameter varies with $a$. Different kinds of lines
represent a series of values of $\beta=-0.5, -0.1, 0, 0.1, 0.5$ with
$\Omega_{m0}=0.27$.}
\label{hqa}
\end{figure}

\begin{figure}
\centering
\includegraphics[width=84mm]{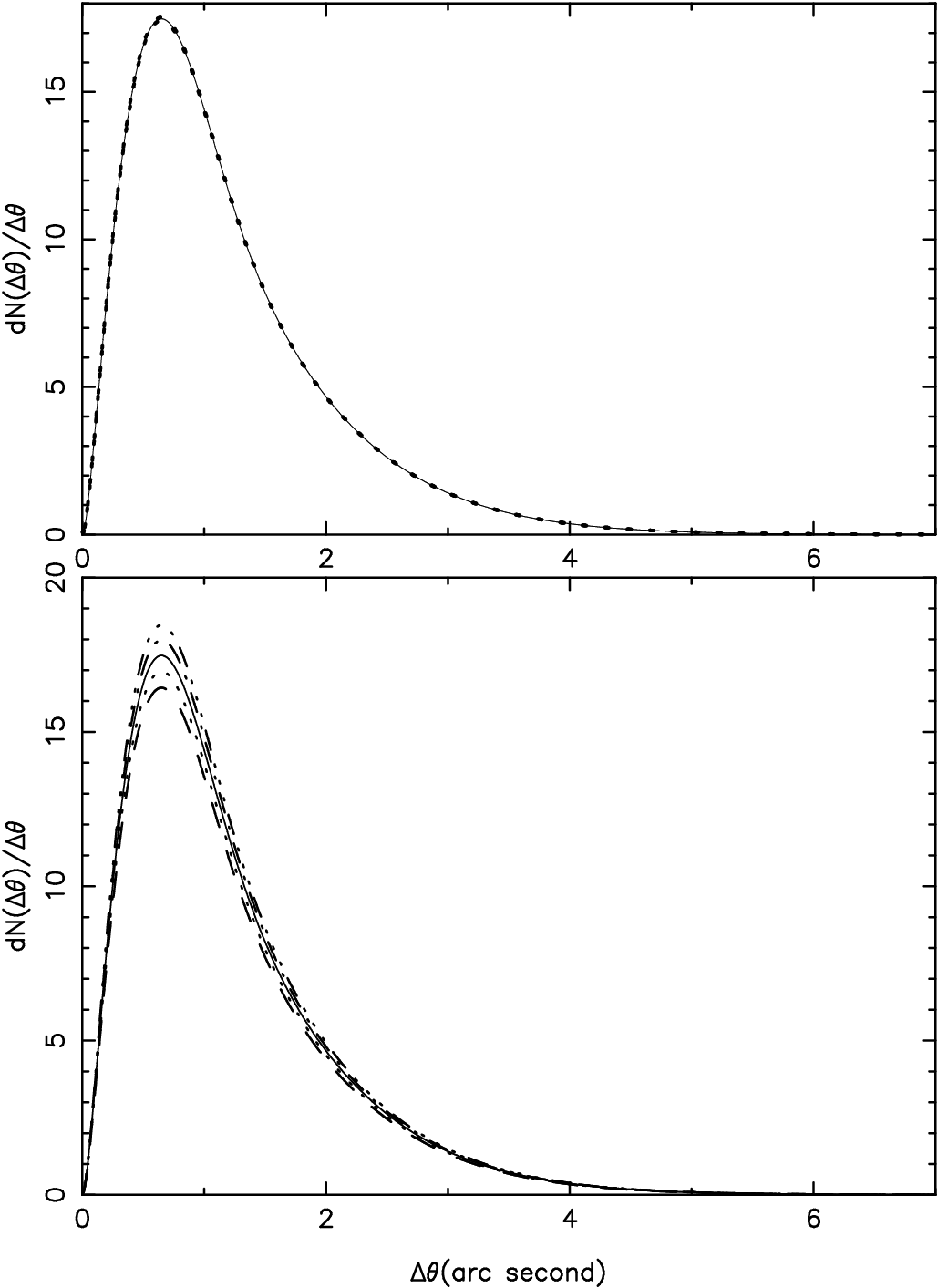}
\caption{The expected number density based on
$f(R)=R-\alpha(-R)^\beta$ for different choices of $\beta$. The
upper panel is for the cases of $\beta=-0.005, 0, 0.005$ and the
lower panel for $\beta=-0.1, -0.05, 0,0.05 ,0.1$ from top to down.
In all cases, $\Omega_{m0}=0.27$. The case $\beta=0$ corresponds to
$\Lambda$ CDM model.}
\label{dn1}
\end{figure}

\subsection{the background evolution}
 Before using  Eqs.(\ref{ar}) and (\ref{h2r}) to determine the
cosmological evolution behaviour, we must give the initial
conditions: ($\rho_{m0}$, $H_0$, $R_0$), it is noting that a
subscript 0 denotes evaluation  at the present time. By putting
$f(R_0)$ into  Eq.(\ref{h2r}), we find that the combination $H_0/R_0$ appears as
a single quantity in the equation, in turn, we choose $H_0=1$ to
compute $R_0$. Hence, from $a_0=1$ and the obtained
value of $R_0$, the values of $\rho_{m0}$  and
$\Omega_{m0}=\kappa\rho_{m0}/(3{H^2_0})$ are fixed. In other words, amongst
the variables $\alpha$, $\beta$ and $\Omega_{m0}$, two of them are
independent. So far, we have obtained all the initial quantities.

Now let us use the equations we derived to examine the
evolution history of the universe at late-time. With the present matter
component
value of $\Omega_{m0}=0.27$, the changing of the Hubble parameter and
scale factor with the curvature are plotted in Figure \ref{ahr}. It is
easy to see that for any choice of $\beta$, the smaller the curvature $|R|$, the
larger the scale factor, and the smaller the rate of change of $a$ ($a'(R)$).
Contrarily, the Hubble parameter $H$ drops with decreasing $|R|$. Therefore, it
is worth noting that the above mentioned
constraint conditions for further restricting the parameter
$(\alpha, \beta)$ space are satisfied as long as the present
curvature $R_0$ meet these conditions. And curvature
$R_0$ decreases with increasing $\beta$.

The background dynamical behaviour is well described by lines
plotted in figure \ref{hqa}. Obviously, the Hubble parameter drop
sharply with the expanding universe, at the same time, the value of
deceleration parameter  is changed from the positive to negative,
representing that the universe evolves from deceleration to
acceleration. Different choice of $\beta$ shows different evolution
history. At any given time, the larger value of $\beta$ harmonizes
the larger Hubble parameter and, the larger deceleration parameter
after a certain time. It is emphasized that larger value of $\beta$
corresponds larger acceleration for our accelerated expanding
universe at present time. Furthermore, with $\beta$ increasing, the
time for the transition from the gravity-dominated era to the
DE-dominated era is gradually late.

\begin{figure}
\centering
\includegraphics[angle=270,width=84mm]{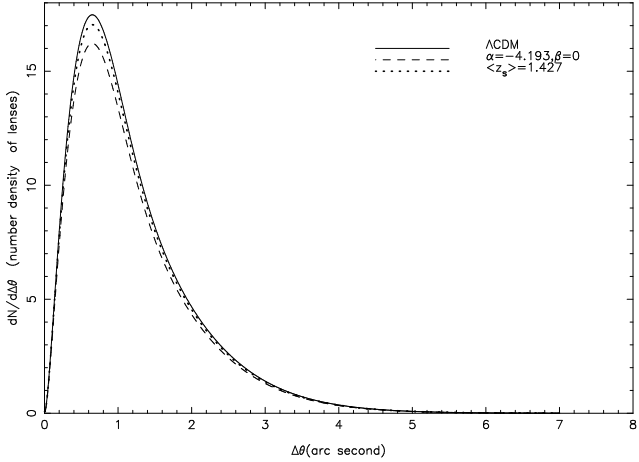}
\caption{The expected number density in the present model for
$\Lambda$CDM, our best-fitted standard model with the value of
$(\alpha,\beta)=(-4.193,0)$ and the $\Lambda$CDM in the case of
$\langle z_s\rangle=1.427$ respectively. The two former number
density lines based on the bin-method.} \label{dn2}
\end{figure}

\begin{figure}
\centering
\includegraphics[angle=270,width=84mm]{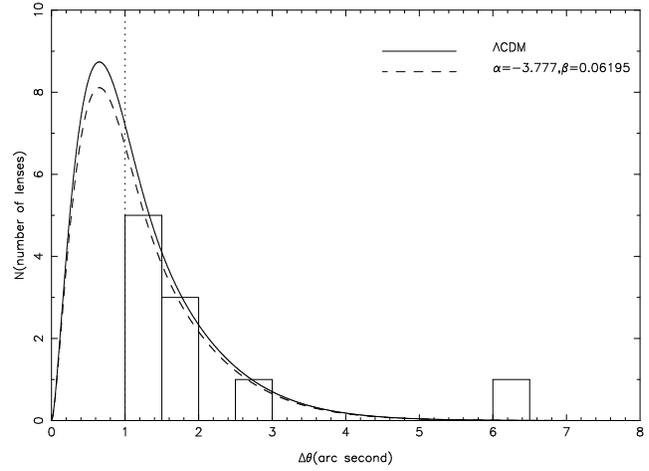}
\caption{The lensed source quasar number as a function of image
separation $\Delta\theta$. The histogram is the number distribution
of the SQLS DR3 statistical lens sample, in which there is a
cluster-scale lens with image separation $14.62''$. We rule out it
because our statistical probability is computed based on the SIS
modeled galaxy-scale lenses. The bin-size is $0.5{''}$, so the lines
stand for $N(\Delta\theta)=\frac{dN}{d\Delta\theta}\cdot0.5{''}$,
representing the lens number around $\Delta\theta$ with width of
$0.5{''}$. The solid line shows $\Lambda$CDM model, while the dashed
line stands for our best-fitting results $(\alpha, \beta)=(-3.777,
0.06195)$. The dotted line at $\Delta\theta=1{''}$ indicates the
SQLS resolution limit.} \label{n}
\end{figure}

\subsection{the SQLS sample constraints}
In this section, we consider the constraints arising from the strong
lensing observations on our gravity model. We have calculated the
number density of lensed quasars of the present model using the SQLS
sample. Specifically, from the differential lensing probability
$P(\Delta\theta, R_s)$, which is multiplying likelihood with respect
to two variables, image separation $\Delta\theta$ and $R_s$ (determined by the
redshift $z_s$ of the source quasars), if we know the  number
distribution $N(z_s)$, then the expected number
density is given by
\begin{equation}
dN(\Delta\theta)/d\Delta\theta=\int{N(R_s)P(R_s,\Delta\theta)dR_s},
\end{equation}
and the expect lens number with image separations between
$\Delta\theta_1''$ and $\Delta\theta_2''$ is computed by
\begin{equation}
N=\int^{\Delta\theta_2}_{\Delta\theta_1}{\frac{dN(\Delta\theta)}{d\Delta\theta}}
d\Delta\theta,
\end{equation}
which can be used to compare with 11 lenses in the sample to constrain the
parameters of the $f(R)$ model. Furthermore, it is important to
point out that, according to the integral mean value theorem, we can also
rewrite the expected lens number with image
separation ranging from $\Delta\theta_1{''}$ to $\Delta\theta_2{''}$
 as follows:
\begin{equation}
N=\frac{dN(\Delta\theta)}{d\Delta\theta}(\Delta\theta_2-\Delta\theta_1),
\end{equation}
where $\Delta\theta$ is one exist value between $\Delta\theta_1$ and
$\Delta\theta_2$.

The sample we used has 22683 quasars, and the redshifts range from
0.6 to 2.2. To get $N(z_s)$, we count the source quasar number for
each redshift bin, with bin-size $\Delta z_s=0.05$. We investigate
how the strong lensing probability depends on the $f(R)$ model we
choose. In Figure \ref{dn1}, we present the lens number density
based on SQLS lens sample with a series of choices of $\beta$. When
$\beta$ is small enough, for example $\beta=\pm0.005$, the number
density is hard to distinguish from $\Lambda$CDM  ($\beta=0$).
Inevitably, we can conclude that the strong lensing statistics
restricts the parameter $\beta$ to not beyond the order of
$10^{-3}$. For any choice of $\beta$, the number density sharply
increases with  small image separations, and is slowly down with
large image separations. At about $\Delta\theta=5{''}$, the lines
drop to zero. Different choices of the parameter $\beta$
significantly influence the curves near the peak, and $\beta$
increases while the peak declines, but they cannot change the
position of the peak, which is located approximately at
$\Delta\theta=0.6{''}$ and is beyond the image separation range of
SQLS lens sample.

We derive constraint on the parameter $\alpha$ assuming $\beta=0$.
The best-fit value is $\alpha=-4.193$, so we get
$\Omega_{m0}=0.301$. From $\Delta\chi^2=1$ we can further get the
95\% confidence interval for $\alpha$ that is [-4.633, -3.754].
Above approach to get the lens number density can be called
bin-method. As another interesting approach, we assume that the
redshifts of the source quasars follow a Gaussian distribution,
which is used to fit the redshift distribution of 22683 quasars. The
best fit gives the mean $\langle z_s\rangle=1.427$ and the
dispersion $\sigma=0.519$. In this approach, The predicted number
density is computed by $N(\Delta\theta)=22683\times P(\Delta\theta,
\langle z_s\rangle)$. Which is also plotted in Figure \ref{dn2}, in
order to compare the cases of, standard $\Lambda$CDM and our
obtained best-fitting standard model based on bin-method.

\begin{figure}
\centering
\includegraphics[angle=270,width=84mm]{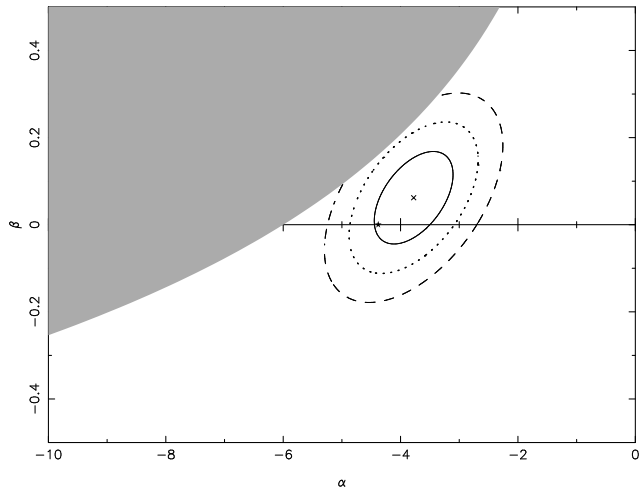}
\caption{The 68.7\%, 95.7\%, 99.7\% joint confidence contour
estimated from $\Delta\chi^2=2.3, 6.17$ and $11.8 $ on the
$(\alpha,\beta)$ plane, arising from fitting the strong lensing
sample. Our best fit value is $(\alpha,\beta)=(-3.777,0.06195)$,
which is marked with a cross, meanwhile the $\Lambda$CDM model
$(\alpha,\beta)=(-4.38,0)$ is marked with a star.}
\label{contour1}
\end{figure}

\begin{figure}
\centering
\includegraphics[angle=270,width=84mm]{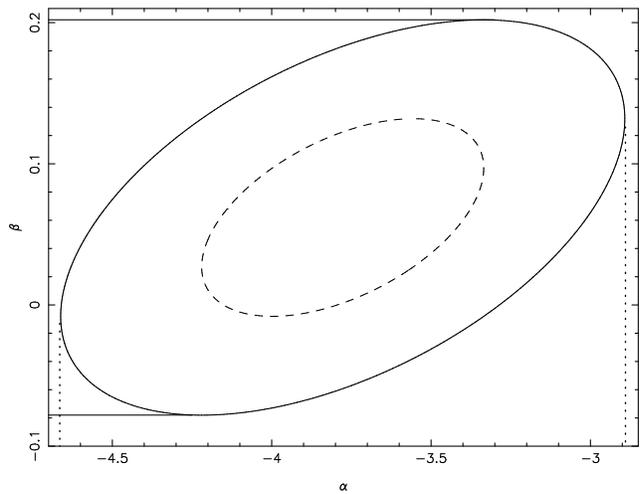}
\caption{The confidence contour plotted from $\Delta\chi^2=1.0, 4.0$
to estimate the region for the single parameter $\alpha$ or $\beta$.
The projections on the vertical and horizontal axises are
corresponding to $1\sigma$ and $2\sigma$ confident regions for the
parameter $\beta$ and $\alpha$, respectively. The solid line and
dotted line show the tangents to the ellipse representing 95\%
confidence level, parallelling to the horizontal axis and vertical
axis respectively.} \label{contour2}
\end{figure}

As mentioned earlier, the typical type of $f(R)$ gravity theories in
form of $f(R)=R-\alpha(-R)^\beta$ have been extensively studied by
using all aspects of observation data. \citet{Amarzguioui2006}
combined the SNIa data from \citet{Riess2004} with the baryon
acoustic oscillation length scale data \citep{Eisenstein2005} and
the CMB shift parameter \citep{Spergel2003} to constrain the
parameters of above gravity model, and found that the best-fit model
with the value is $(\alpha, \beta)=(-3.6, -0.09)$. Similarly, by
using the supernovae data by the supernova Legacy Survey
\citep{Astier2006} together with the baryon acoustic oscillation
peak in the SDSS luminosity red galaxy sample \citep{Eisenstein2005}
and the CMB shift parameter \citep{Spergel2003,Spergel2007},
\citet{Fay2007} constrained the parameters to $\alpha=-4.63$ and
 $\beta=0.027$ in the best-fit case. Both works found that the
 $\Lambda$CDM model is well within 68.3\% confidence level and the constrained
parameter $\beta$ is in $10^{-1}$ magnitude.
 \citet{Koivisto2006} calculated the matter power spectrum and matched it with
 the measurements of SDSS \citep{Tegmark2004b}. The result is that the favored
values of
 $\beta$ are in order of $10^{-5}$. Meanwhile, the matter power spectrum
 is also calculated with TT CMB by \citet{Li B.2007-2}. Where,
 the combined data from WMAP \citep{Page2007,Hinshaw2007}, SNLS and SDSS
\citep{Tegmark2004b} is used to restrict the model parameters and it is found
that $\beta$ is confined in a even
 smaller order of $10^{-6}$. From this point, we can see that the
 matter power spectrum is rather sensitive to the model parameter
 $\beta$, and the model obtained from the observational constraints on the
 mater power spectrum is seeming indistinguishable from the standard
 one with $\beta=0$. Here, we extend these studying by calculating the strong
lensing probability and using the SQLS DR3 lens sample to constrain
the parameters. In figure \ref{n}, we have plotted $N(\Delta\theta)$
for the $\Lambda$CDM model and our best-fitting model with
$\alpha=-3.777$ and $\beta=0.06195$, as well as the histogram from
the SQLS DR3 statistical sample. Except for both small and large
image-separation ends, the non-linear gravity model predicts less
number of lenses than those predicted by the $\Lambda$CDM model,
which is linear in the form of Ricci scale R. The cosmological
parameters $\Omega_{m0}=0.285$ and $q_0=-0.544$ are derived from the
best-fit case, which are broadly in agreement with other
measurements. We also present the contours for the joint
distribution of $\alpha$ and $\beta$ in Figure \ref{contour1}, we
can see that the $\Lambda$CDM model lies in the 68.3\% confidence
contour, which is consistent with the results of other works. Figure
\ref{contour2} shows
 the $1\sigma$ and $2\sigma$ confidence region of the single parameter
$\alpha$ or $\beta$ from the projection of the joint confidence
lever estimated from $\Delta\chi^2=1.0$ and $\Delta\chi^2=4.0$. We
have found that the  model $f(R)=R-\alpha(-R)^\beta$ is compatible
with the lensing observational data subject to the parameter
constraints [-4.67, -2.89] and [-0.078, 0.202] for $\alpha$ and
$\beta$, respectively, at the 95\% confidence level . Obviously, the
data coming from the SQLS DR3 lens sample can only  confine the
parameter $\beta$ to order of $10^{-1}$.

\section{discussion and conclusions}

To summarize, in this work we have derived new constraints on
$f(R)=R-\alpha(-R)^\beta$ in the Palatini formalism, using the
statistics of strong gravitational lensing and the lens sample come
from SQLS DR3. We use  $f(R)$ within Palatini approach because of it
being free from the instabilities, passing the solar system and
having the correct Newtonian limit. This typical type of $f(R)$
theories in form of $f(R)=R-\alpha(-R)^\beta$ have been recently put
forward to explain the late-time expansion phase dominated by dark
energy and have been well studied by using the cosmology
measurements including background universe evolution parameters,
matter power spectrum and CMB. In our paper, we assume that the
$f(R)$ gravity theories have hardly detectable effect on the gravitational 
lensing phenomena of galaxy-scale halos, and then the standard lensing is a 
realistic approximation of lensing in Palatini $f(R)$ gravity theories.
It is worth emphasizing that in the SQlS lens
sample, there is a cluster-scale lens, we use SIS model to calculate
the differential lensing probability to fit the other 10 lens data
to avoid destroying the well-defined statistical lens sample.

Considering the FRW setting, we have shown the background evolution
for the late expansion era of matter-dominated phase based on our
$f(R)$ theory. The expected lens number density arising from the
SQLS DR3 sample is not as sensitive as the matter power spectrum to
the parameter $\beta$.  The number density distributions for
different values of $\beta$ are indistinguishable to $10^{-3}$ in
magnitude, while the matter power spectrum changes a lot even if
$\beta$ varies only by a tiny amount, such as an order of  magnitude
$10^{-5}$ \citep{Li B.2007-2, Koivisto2006}. Therefore, the matter
spectrum can constrain $\beta$ to a small region and make the model
hardly deviation from the $\Lambda$CDM model. But the values of our
best fit model, $\alpha=-3.777$ and $\beta=0.06195$, are not much
more different from the standard one with $\alpha=-4.38$ and
$\beta=0$, implying that the model studied is not significantly
preferred over the standard model. The results are consistent with
the works of \citet{Amarzguioui2006}, \citet{Fay2007} and \citet{Carvalho2008}, in which
the best-fitting models are $(\alpha, \beta)=(-3.6, 0.09)$, $(\alpha, \beta)=(-4.63, 0.027)$ and $(\alpha,
\beta)=(-4.7, 0.03)$, respectively, and we all found that
$\Lambda$CDM lies in the 68.3\% confidence level. Finally, we
confine the allowed values of $\alpha$ and $\beta$ to the ranges
[-4.67,-2.89] and [-0.078, 0.202] at the 95\% confidence level, and
conclude that the SQLS DR3 statistical lens sample can only
constrain the parameter $\beta$ to $10^{-1}$ in magnitude.

In our expected number density distribution (Figure \ref{dn1}), for
any value of  $\beta$ with the fixed $\Omega_{m0}=0.27$, the peak is
at about $\Delta\theta=0.6{''}$. But the lens sample we used, with
the image-separation ranging from $1{''}$ to $6.17{''}$, provides no
data in $\Delta\theta<1{''}$, this significantly restrict the
constraint precision of the parameters. The CLASS radio lens sample
has the lens image separations of $0.3{''}<\Delta\theta<3{''}$.
Naturally, jointing the CLASS and SQLS samples to constrain the
model  will give better results. Of course, a more larger lens
sample from future surveys, \cite[e.g., Square Kilometer
Array,][]{Koopmans2004}, will be able to make the constraints on the
model more stringent. Furthermore, using the SIS model and NFW model
together to constrain the $f(R)$ theories models arising from the
lensing observation data containing both galaxy lens and cluster
lens is future work.

Finally, we emphasize that our paper is the first try to test $f(R)$
gravity theories using strong lensing statistics. The reason may be
that, the possible deviation of gravitational lensing based on
$f(R)$  gravity theories from that built on GR is not well studied
practically. Our results hold on the valid hypotheses that this possible 
deviation is hardly detectable, and they are not contrary to other works.
As mentioned in the introduction, if the $f(R)$ 
theories are successful models to explain
the nature of DE,  they should have no important effects which can
be detected on the galaxy-scale and cluster-scale dark halos. The
reason is that DE is introduced to explain the acceleration of the
universe, which is, of course, in the cosmological scale. Namely, DE
smoothly permeates the regions of smaller scales, such as galaxies
and clusters, then it has no contributions to the inhomogeneity of
gravitational potential for such halos. For the dark matter as the
dominator of the gravity field, the SIS and NFW lens models are well
consistent with the observations. It is therefore necessary to test
the $f(R)$ theories through verifying that the phenomenology is not
contradict those observational results that do agree with the
predictions of the standard lensing theory. This critical testing
will be presented in future work.

\section*{Acknowledgments}
We warmly thank Masamune Oguri for useful discussion and providing
the redshift data of the source quasars. We also thank P.J. Zhang
for his helpful comments on the manuscript and revision of the
abstract and the anonymous referee for very useful comments that
improved the presentation of the paper. XJY also appreciate all
members in cosmology group in NAOC for their help. This work was
supported by the National Natural Science Foundation of China under
grant 10673012, CAS under grant KJCX3-SYW-N2 and National Basic Research
Program of China  (973 Program) under grant No.2009CB24901.

\end{document}